\documentclass[aps,prd,twocolumn,showpacs,floatfix]{revtex4-1}
\usepackage[utf8]{inputenc}
\usepackage[OT4]{fontenc}
\usepackage{amsmath}
\usepackage{amsfonts}
\usepackage{amssymb}
\usepackage{axodraw}
\usepackage{graphicx}
\begin{document}
\title{Scalar isovector resonance photoproduction through the final state meson-meson interactions}
\author{\L{}ukasz Bibrzycki}
\email{lukasz.bibrzycki@ifj.edu.pl}
\affiliation{Chair of Computer Science and Computational Methods, \\Pedagogical University of Cracow, 
Podchor\k{a}\.zych 2, 30-084 Krak\'ow, Poland}
\author{Robert Kami\'nski}
\affiliation{The Henryk Niewodnicza\'nski Institute of Nuclear Physics Polish Academy of Sciences,
Radzikowskiego 152, 31-342 Krak\'ow, Poland}
\begin{abstract}
We construct the amplitudes of $\pi\eta$ photoproduction taking into account
the effects of the $\pi\eta$ - $K\bar K$ interchannel coupling.
The idea of our model is close to the molecular description of scalar
resonances with exception that apart from the pseudoscalar loops we include
also vector mesons in the intermediate state loops. These
amplitudes are used to calculate the $S$-wave cross sections and
mass distributions in the $\pi\eta$ effective mass region
corresponding to the scalar resonances $a_0(980)$ and $a_0(1450)$.
The values we obtained for $a_0(980)$ are comparable with predictions
of other models while the cross section for $a_0(1450)$ is about an
order of magnitude larger than prediction based on the quark model. We show
that the amplitudes with loops containing vector
mesons calculated in the on-shell approximation are not suppressed
in contrast to amplitudes containing only pseudoscalar loops. We estimate
the cross sections for the $P$-  and $D$- waves in the $\pi\eta$ channel.
\end{abstract}
\maketitle
\section{Introduction}
With the advent of abundant data already provided or foreseen in near future by experiments like CLAS12 and GlueX at 
Jefferson Laboratory, BESIII at BEPCII and COMPASS at CERN, new perspectives open
in the spectroscopy of light quark mesons with particular stress put on studying their internal structure and 
production mechanisms. Apart from the $\pi\pi$ and $K\overline{K}$ channels which traditionally evoked the interest 
in the context of studying the properties of scalar resonances, the $\pi\eta$, $3\pi$, $\eta\pi\pi$ and 
$\pi\eta'$ channels will be observed in photoproduction, electroproduction, pionproduction and charmonia decays, 
with statistics unreachable so far. 

New data are necessary to address the long standing questions of the internal structure of both isoscalar and 
isovector scalar resonances. The common belief is that at least the lightest scalars like $f_0(500)$, $f_0(980)$ and 
$a_0(980)$ are not ordinary $q\overline{q}$ states
but rather the tetraquarks or mesonic molecules \cite{Baru:2003qq}. In this respect, the radiative decays, where 
scalar mesons are either among decay products or decay themselves
are of special interest. In various models of the scalar meson structure the photonic probes couple either to quarks 
or to underlying meson loops. The observables constructed to describe these processes provide the means to
discriminate among the models \cite{Achasov:1987ts,Close:1992ay}. The photoproduction can be treated as a process 
providing information complementary to that obtained from radiative decays.

The photoproduction of scalar and tensor resonances in $K\overline{K}$ and $\pi\pi$ channels has been discussed 
in articles \cite{JiKaLe,BibLesSzcz,BibLes,BiKa}. Here, we 
are interested in the $\pi\eta$ channel which in the $S-$wave is strongly influenced by isovector resonances
$a_0(980)$ and $a_0(1450)$. The $\gamma p\to\pi^0\eta p$ reaction has been recently studied at energies below 1.4 
GeV \cite{Kashevarov:2009ww,Kaser:2015bum,Annand:2015fea} using the TAPS and Crystal Ball detectors at MAMI. 
Theoretical tools 
had been developed to describe the polarization observables measured at these low energy $\pi\eta$ photoproduction 
experiments \cite{Doring:2010fw,Fix:2010nv,Fix:2013hta}. Near threshold photoproduction of pseudoscalar pairs is
dominated by excitation of baryonic resonances in the intermediate state. It was shown, however, in \cite{MaOsTo} 
that pions emitted in $\Delta\to\pi N$ 
and $\gamma N\to\Delta\pi$ vertices do not combine into the $S-$wave, thus do not participate in the scalar meson 
photoproduction. This argument also applies to pions emitted in the $N^\ast(1529)\to\Delta\pi$, $\Delta\to\pi N$ 
cascade. Contrary to that, the emission of two pions (or kaons) from a nucleon line with photons attached to pion 
lines (plus contact terms) is claimed to give the major $S-$wave contribution. Mesons $f_0(980)$ and $a_0(980)$ are 
then generated by iteration of the pseudoscalar loops using the lowest order chiral lagrangian vertex.
Application of the chiral methods makes this approach limited to low energies, whereas in this work we are 
interested in the high energy $\pi\eta$ photoproduction 
corresponding to photon energies around 10 GeV. Theoretical analysis of the $\pi\eta$ photoproduction at this photon 
energy range is timely and important as this is the energy designed for new experiments at JLab - CLAS12 and 
GlueX.

According to Regge phenomenology, the high energy regime of the photoproduction is dominated by the $t$ channel 
Reggeon exchanges. Such reggeised amplitudes were used to describe the photoproduction of scalar mesons in 
\cite{DonKal}. Light scalars were treated there as molecules or tetraquarks or members of conventional 
$q\overline{q}$ ground state nonet, while various scenarios of mixing between isoscalars and scalar glueball were 
tested. In the case where $a_0(980)$ (and $f_0(980)$) photoproduction is due to the photon coupling to kaon (and pion) 
loops, the respective cross sections were found to be small as suggested by the large $N_c$ considerations 
\cite{DonKal,OlOs}.
Amplitudes with $t$ channel vector meson exchange supplemented with the final state pseudoscalar pair interaction 
were used to describe the $f_0(980)$ and $a_0(980)$  photoproduction in the $K^+K^-$ channel
 \cite{JiKaLe,BibLesSzcz} as well as the $f_0(980)$ and $f_2(1270)$ photoproduction in the  $\pi^+\pi^-$ channel
 \cite{BibLes,BiKa}.
In spite of the fact that these models also engage meson loops, the calculated cross sections were found to be of 
the size similar to those obtained with quark and chiral models. This can be explained by the fact that 
suppression of the amplitudes observed in models which employ the iteration of kaon (and pion) loops, does not 
apply to amplitudes where
vector mesons like $\rho$, $\omega$ and $K^\ast$ enter the loop. This is precisely the case discussed in 
\cite{JiKaLe,BibLesSzcz,BibLes,BiKa}.

In this paper we propose the description of the photoproduction of scalar isovector resonances as a final
state interaction effect. In particular we analyze the effect of the inclusion of vector mesons in 
the intermediate state loops as compared to previous analyses based only on pseudoscalar loops. 
Our approach is additionally justified by the common belief that the $a_0(980)$ wave function may contain 
a dominant molecular component. 
In fact we extend our model to $a_0(1450)$ photoproduction. In this case, however, our approach is 
rather phenomenological and based on the fact that the unitary amplitudes we use to describe the final state 
interactions include both $a_0(980)$ and $a_0(1450)$.

The structure of paper is as follows. General description of the model is given in section \ref{Struct}. 
In sections \ref{Born} and \ref{FSI} we collect the main components of the 
formalism used to describe the initial meson pair photoproduction and the final state interactions. 
Most of formulas contained in these two sections were published in separate papers but we collect them to 
make the paper self-contained. Section \ref{Results} contains the numerical results 
obtained with the model for the $S-$wave and for higher partial waves. In section \ref{Summary} we discuss our 
results, limitations of the model and directions of 
the future study.
\section{Structure of the model}\label{Struct}
\begin{figure}[htbp]
\centering
\includegraphics[scale=.7,clip]{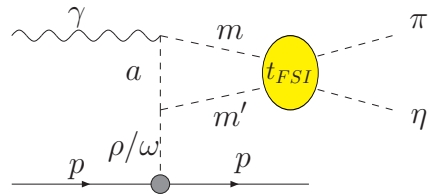}
\caption{Diagram for the photoproduction process with final state interactions}
\label{diag-fsi}
\end{figure}

The dominating decay channels of $a_0(980)$ and $a_0(1450)$ are $\pi\eta$ and $K\overline{K}$. Thus constructing the 
amplitudes of the resonant $\pi\eta$ photoproduction is inevitably the coupled channel problem 
whose solution must take into account the $K\overline{K}$ intermediate states.
The model described in this paper assumes that the isovector resonances  observed in the 
$\pi\eta$ channel emerge as an effect of the $\pi\eta$ and $K\overline{K}$ interactions in the final state. 
Consequently, the resonance
photoproduction is described as a two stage process. In the first stage a pair of pseudoscalar mesons $\pi\eta$, 
$K^0\overline{K^0}$ or $K^+K^-$ is photoproduced. This stage is described in terms of the Born amplitudes. Then
the meson pair undergoes the final state interactions described in terms of the coupled channel and unitary 
amplitudes. The Born amplitude originally contains the contributions of all partial waves. Therefore we need
to project the whole amplitude on the $S-$wave in order to proceed with the analysis of scalar resonances.
In view of the general
partial wave analysis this procedure can be most conveniently performed in the rest frame of the resonantly 
interacting system.
Consequently we perform our calculations in the helicity system which is the CM system of the interacting $\pi\eta$ 
($K\overline{K}$) pair, where the $z$ axis is directed opposite to the momentum of the recoil proton and $y$ axis is 
perpendicular to the production plane determined by momentum vectors of the incoming photon and target proton. 
Then the $x$ 
axis is defined by the relation $\hat{x}=\hat{y}\times\hat{z}$.
The diagram representing a complete photoproduction amplitude is shown in Fig.\ref{diag-fsi}, where $a$, 
$m$ and $m'$ correspond to mesons exchanged in the intermediate state. 
The general form of the $\pi\eta$ photoproduction amplitude, which takes into account the final state interactions 
is
\begin{equation}
\begin{split}
&\langle \lambda' M |A_{\pi\eta}|\lambda_\gamma \lambda \rangle=
    \langle \lambda' M|V_{\pi\eta}|\lambda_\gamma \lambda\rangle+\\&
    4\pi\!\!\sum_{\{mm'\}}\!\int_0^\infty\!\!\frac{{k'}^2 dk'}{(2\pi)^3}F(k') 
    t_{\pi\eta;mm'} G_{mm'}
    \langle \lambda' M|V_{mm'}|\lambda_\gamma \lambda\rangle
\end{split}
\label{photoamp}
\end{equation}
where $V_{\pi\eta}$ ($V_{mm'}$) is the Born amplitude of the $\pi\eta$ ($mm'$) pair photoproduction, 
$t_{\pi\eta;mm'}$ is the coupled channel
 scattering amplitude, $\lambda, \lambda', \lambda_\gamma$ and $M$ are respectively the helicities 
of the initial and final proton, photon helicity and projection of
the $\pi\eta$ system angular momentum on the spin quantisation axis $z$, which is equivalent to the helicity of the
dynamically created resonance. $G_{mm'}$ is the propagator of the intermediate $mm'$ pair defined as
\begin{equation}
G_{mm'}=\frac{-1}{M_{mm'}-{M'}_{mm'}(k')+i\epsilon}
\end{equation}
and $F(k')$ is the form-factor needed to regularize the divergent 
mesonic loop of the diagram shown in Fig.\ref{diag-fsi}. The form factor depends on $k'$ i.e. the momentum in the 
intermediate loop. The summation in Eq.(\ref{photoamp})
 runs over intermediate $\pi\eta$ and $K\overline{K}$ states. As shown in \cite{JiKaLe} the off-shell part of the 
photoproduction amplitude is rather strongly dependent on the value of the cut-off parameter. However, fixing the 
cut-off without additional experimental input is rather difficult task. 
On the other hand, the dominance of the on-shell effects was previously conjectured in \cite{MaOsTo} (which suggests 
rather small value of the cut-off parameter).
This is why in the present calculations we limit ourselves to the on-shell part of the amplitude. 
We stress that Eq.(\ref{photoamp}) in principle enables the computation of the photoproduction amplitudes for any 
partial wave. Here, we are mainly concerned with the $S-$wave. This is because the experimental situation in the 
$\pi\eta$ channel is much worse than in the 
$\pi\pi$ channel, to the extent that no experimental phase shifts or inelasticities are accessible. So, one is left 
with models of the $\pi\eta$ scattering \cite{Oller:1998hw,GomezNicola:2001as,Black:1999dx}. These models are
generally concerned with the $S-$wave and employ 
the chiral perturbation theory supplemented with constraints of unitarity, analyticity and crossing symmetry. Their 
application is limited by the $\pi\eta$ energies corresponding to the mass of the $a_0(980)$ resonance. 
As we want to describe also the photoproduction of the $a_0(1450)$ resonance, we use the 
$S-$wave $\pi\eta\to\pi\eta$ and
$K\overline{K}\to\pi\eta$ amplitudes described in \cite{Lesniak:1996qx,Furman:2002cg}.

Finally, after momentum integration in Eq.(\ref{photoamp}) (dropping off the off-shell part) and projection on the 
$S-$wave, we arrive at the following coupled channel $S-$wave $\pi\eta$ photoproduction amplitude (we omit spin 
indices for simplicity):
\begin{multline}
A_{\pi\eta}=\left[1+i r_{\pi\eta}t_{\pi\eta}^{I=1}\right]
V_{\pi\eta}+\\
r_{K\overline{K}} t_{\pi\eta;K\overline{K}}^{I=1}\frac{1}{\sqrt{2}}\left(V_{K^+K^-}+V_{K^0\overline{K^0}}\right),
\label{ampls}
\end{multline}
where $r_{\pi\eta}=-k_{\pi\eta} M_{\pi\eta}/8\pi$, $r_{K\overline{K}}=-k_{K\overline{K}}M_{\pi\eta}/8\pi$,
 $M_{\pi\eta}$ is effective mass of the $\pi\eta$ system, $t_{\pi\eta}^{I=1}$
and $t_{K\overline{K};\pi\eta}^{I=1}$ are isovector elastic $\pi\eta$ scattering amplitude and 
$K\overline{K}\to\pi\eta$ transition amplitude respectively. The second term in Eq.
\ref{ampls} contains the isovector combination of the Born amplitudes.
$k_{\pi\eta}$ and $k_{K\overline{K}}$ are CM momenta in the $\pi\eta$ and $K\overline{K}$ channels defined as
\begin{equation}
k_{\pi\eta}=\frac{\sqrt{[M_{\pi\eta}^2-(m_\eta+m_\pi)^2][M_{\pi\eta}^2-(m_\eta-m_\pi)^2]}}{2M_{\pi\eta}}
\end{equation}
and
\begin{equation}
k_{K\overline{K}}=\sqrt{\frac{M_{\pi\eta}^2}{4}-m_K^2}.
\end{equation}
In what follows we use numerical channel indices for the elements of the final state scattering amplitude and denote
$t_{\pi\eta}$ as $T_{11}$ and $t_{\pi\eta;K\overline{K}}$ as $T_{12}$ respectively.
\section{Born amplitudes}\label{Born}
While constructing the Born amplitudes of the $\pi\eta$, $K^+K^-$ and $K^0\overline{K^0}$ photoproduction we 
follow the approach developed for the $S-$wave photoproduction in \cite{JiKaLe} and generalized for higher partial 
waves in \cite{BiKa}.
The idea is to start from a set of Feynman diagrams relevant for a given process and calculate the basic Born 
amplitudes. Then these Born amplitudes are projected on the partial wave of interest in the s-channel helicity 
system.
For high energy applications the reggeised versions of the partial wave Born amplitudes are used.
It is worth of mentioning that the same set of model parameters applies to any Born partial wave amplitude which
makes the approach very economical.  The parameters, like coupling 
constants and form factor range parameters 
are taken from the Bonn model \cite{Bonn} or calculated from radiative decay widths, while strong meson couplings 
are derived from the $SU(3)$ symmetry relations (see \cite{JiKaLe} for a list of parameter values). So, if we 
neglect the off-shell contributions, then the model does not engage any new parameters apart from those already
used in other applications.

\begin{figure*}[ht]
\centering
\includegraphics[scale=.75,clip]{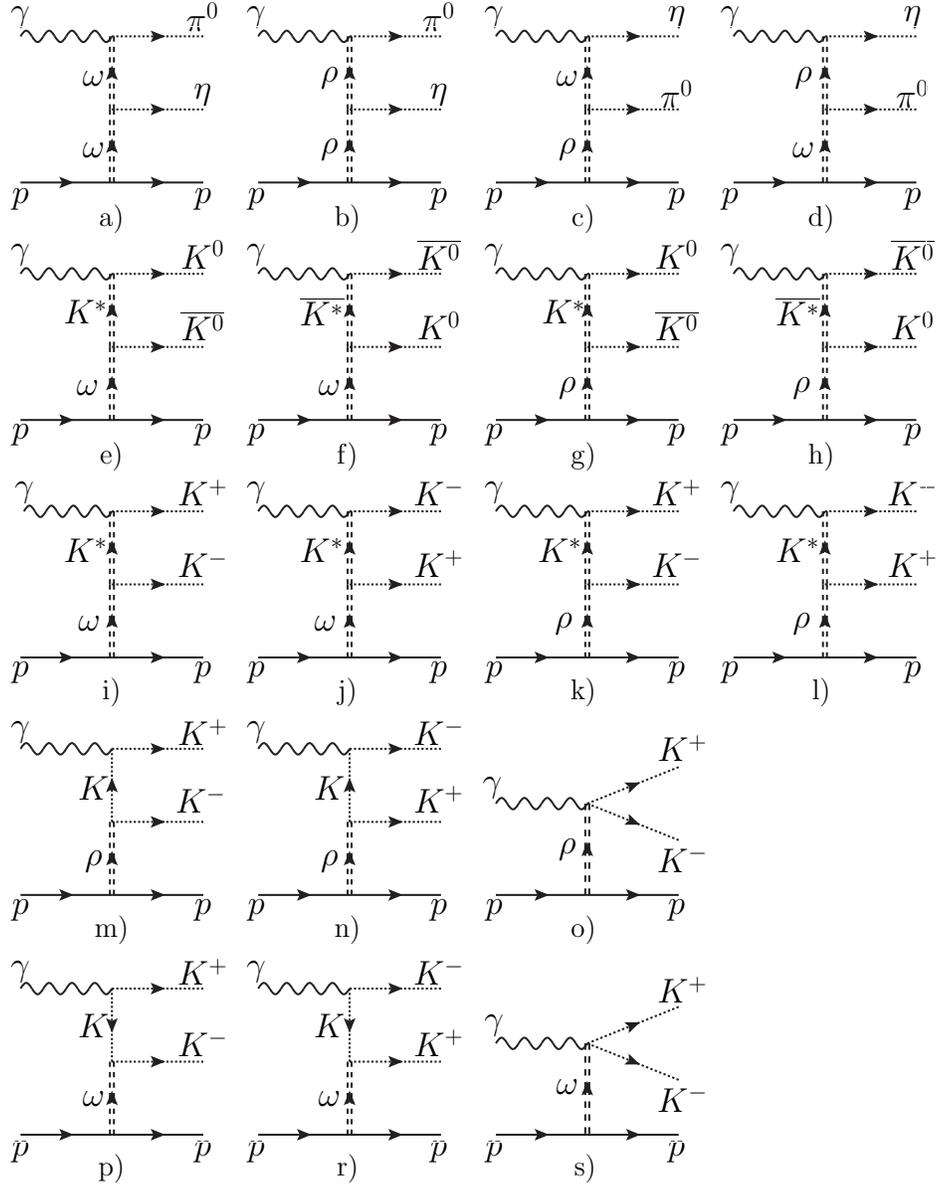}
\caption{Diagrammatic representation of the Born amplitudes for $\pi\eta$, $K^0\overline{K^0}$ and $K^+K^-$ intermediate states.}
\label{FeynmanDiagrams}
\end{figure*}

The set of Feynman diagrams we use to calculate the amplitudes describing the first stage of the reaction is shown in Fig. \ref{FeynmanDiagrams}.
These diagrams are of two types: type I diagrams, where
the vector meson is exchanged in the lower vertical line while the pseudoscalar is exchanged in the upper line and 
type II diagrams, where vector mesons are exchanged in both vertical lines. Type II diagrams are the only ones 
which contribute to 
the neutral intermediate states as the photon does not couple to neutral pseudoscalars. Whereas for charged 
intermediate states both type I and type II diagrams participate. Please note, however, that type I diagrams require 
additional contact diagram to secure the electromagnetic current conservation.

The catalogue of both types of diagrams (r=I and r=II) is shown in Table \ref{tabdia}. The first term in parentheses
denotes the upper vertical line meson and the second term - the lower one (we skip diagrams with external mesonic 
lines related by charge conjugation). In principle the model could include the pseudoscalar exchanges in the
lower line but since we are mostly concerned with high energy photoproduction these exchanges should be suppressed
as suggested by Regge phenomenology.
\begin{table}[h!]
\begin{center}
\begin{tabular}{lcc}
\hline
$m m'$ & r=I & r=II \\ 
\hline
$\pi^0\eta$ &  &  $(\omega,\omega)$,  $(\rho,\rho)$, $(\rho,\omega)$, $(\omega,\rho)$\\
$K^0\overline{K^0}$ & &$(K^\ast,\omega)$, $(K^\ast,\rho)$\\
$K^+K^-$ &$(K,\rho)$, $(K,\omega)$& $(K^\ast,\omega)$, $(K^\ast,\rho)$ \\
\hline
\end{tabular}
\caption{Summary of meson exchanges in Born amplitudes.}
\label{tabdia}
\end{center}
\end{table}
The amplitudes corresponding to diagrams of Fig.\ref{FeynmanDiagrams} have a general form of:
\begin{equation}
 V_{mm'}=\sum_{r=I,II}\overline{u}(p',s')J_{r,mm'}\cdot\varepsilon(q,\lambda^\gamma)u(p,s),
 \label{fullamp}
\end{equation}
where $J_{r,mm'}$ is the hadronic current, $u(p,s)$ and $\overline{u}(p',s')$ - wave functions of the initial and 
final proton respectively and $\varepsilon$ - polarisation 4-vector of the incident photon which reads
\begin{equation}
 \varepsilon(q,\lambda_\gamma)=(0,\boldsymbol{\varepsilon}^{\lambda_{\gamma}}),
\label{polvec}
\end{equation}
where
\begin{equation}
 \boldsymbol{\varepsilon}^{\lambda_{\gamma}}=-\frac{\lambda_\gamma}{\sqrt{2}}(\cos\theta_q,i\lambda_\gamma,\sin\theta_q)
\end{equation}
and $\lambda_\gamma$ is photon helicity.
The photon polarisation vector is transverse to photon momentum:
\begin{equation}
 \boldsymbol{q}=|\boldsymbol{q}|(-\sin\theta_q,0,\cos\theta_q),
\label{photmom}
\end{equation}
and
\begin{equation}
 \cos\theta_q=\frac{E^2-{E'}^2-|\boldsymbol{q}|^2}{2|\boldsymbol{q}||\boldsymbol{p}'|},
\end{equation}
where $p$, $p'$, $q$, $k_1$ and $k_2$ are the 4-momenta of the initial and final proton, photon, pion ($K^+$ or $K^0$) and eta ($K^-$ or $\overline{K^0}$), respectively.
The energies $E$ and $E'$ of the initial and final proton respectively as well as photon energy $|\boldsymbol{q}|$ 
can be expressed in terms of Lorentz invariant quantities
\begin{equation}
 E=\frac{s-m^2+t}{2M_{\pi\eta}}, \qquad E'=\frac{s-m^2-M_{\pi\eta}^2}{2M_{\pi\eta}},
\end{equation}
\begin{equation}
 |\boldsymbol{q}|=\frac{M_{\pi\eta}}{2}-\frac{t}{2M_{\pi\eta}},
\end{equation}
where $s$ is the $\gamma p$ CM energy squared, $t$ is the square of the 4-momentum transfer from the initial photon 
to photoproduced $\pi\eta$ system, $m$ is the proton mass.

The amplitude defined in Eq.(\ref{fullamp}) is then $S$-wave projected using the formula:
\begin{equation}
  V_{mm'}^{00}=\frac{1}{\sqrt{4\pi}}\int d\Omega {Y^0_0}(\Omega) V_{mm'}
 \label{pwe}
\end{equation}
In our frame of reference the momenta of photoproduced pion, $K^+$ and $K^0$ (eta, $K^-$ and $\overline{K^0}$) 
 can be 
expressed in terms of the solid angle $\Omega$, ie. $\boldsymbol{k_1}=-\boldsymbol{k_2}=|k|\hat{\kappa}(\Omega)$. 
Thus $\boldsymbol{k_1}$($\boldsymbol{k_2}$) is the pion, $K^+$ or $K^0$ (eta, $K^-$ or $\overline{K^0}$) momentum 
and $\hat{\kappa}=(\sin\theta \cos\varphi,\sin\theta\sin\varphi,\cos\theta)$.
The general form of the current used in Eq.(\ref{fullamp}) is
\begin{equation}
\begin{split}
 J^{\mu}_{r,mm'}&=(\alpha_{r,mm'}g^{\mu\nu}+k_1^{\mu}\beta_{1r,mm'}^\nu+
k_2^\mu\beta_{2r,mm'}^\nu)\\
&\quad\times\lbrace d_{r,mm'}\gamma_\nu+e_{r,mm'}(p+p')_\nu\rbrace.
\end{split}
\label{current}
\end{equation}
Functions $\alpha_{r,mm'}$, $\beta_{1r,mm'}$ and $\beta_{2r,mm'}$ have different form according to whether the
masses of the produced mesons are equal or different. The case of equal masses where the mesons $m$ and $m'$ are 
charge conjugates of each other was discussed in the Appendix B of \cite{JiKaLe}.
In case of different meson masses these functions have a form:
\begin{multline}
\alpha_{II,mm'}=\frac{1}{m^2_m-m^2_{v_1}-2 q\cdot k_m}\left[ -(q\cdot k_{m'})(q\cdot k_m)\right. \\ \left.-(q\cdot k_m)(k_m\cdot k_{m'})+m_m^2(q\cdot k_{m'})\right],
\label{alfa}
\end{multline}
\begin{equation}
\beta_{1II,mm'}=\frac{q(k_m\cdot k_{m'})-k_m(q\cdot k_{m'})}{m^2_m-m^2_{v_1}-2 q\cdot k_m},
\end{equation}
\begin{equation}
\beta_{2II,mm'}=\frac{k_m(q\cdot k_m)-q(m_m^2-k_m\cdot q)}{m^2_m-m^2_{v_1}-2 q\cdot k_m}.
\label{beta2}
\end{equation}
Subscripts $m$, $m'$ and $\upsilon_1$ in Eqs.(\ref{alfa}-\ref{beta2}) refer to upper and lower pseudoscalar meson 
lines and to upper vector meson line respectively as shown in diagrams a-d.
We have put the subscript $r$=II in Eqs.(\ref{alfa}-\ref{beta2})
because all diagrams corresponding to $\pi\eta$ Born photoproduction are type II diagrams. So, 
in case of Born amplitudes of the $\pi\eta$ photoproduction the summation in Eq. (\ref{fullamp}) is over diagrams 
a-d. Functions $d_{r,mm'}$ and $e_{r,mm'}$ in Eq.(\ref{current}) are defined as
\begin{equation}
 d_{r,mm'}=g_{\gamma\upsilon_1 m}g_{\upsilon_1\upsilon_2 m'}(G^{\upsilon_2}_V +G^{\upsilon_2}_T)
 \Pi_{\upsilon_2}(s,t)F_{\upsilon_2 NN}(t)
 \end{equation}
and
\begin{equation}
e_{r,mm'}=-g_{\gamma \upsilon_1 m}g_{\upsilon_1\upsilon_2 m'}\left(\frac{G^{\upsilon_2}_T}{2m}\right)
\Pi_{\upsilon_2}(s,t)F_{\upsilon_2 NN}(t),
\end{equation}
where $g_{\upsilon_1\upsilon_2 m'}$ and $g_{\gamma \upsilon_1 m}$ are meson strong and electromagnetic couplings,
$G^{\upsilon_2}_V$ and $G^{\upsilon_2}_T$ - vector and tensor couplings of vector mesons to nucleon, 
$\Pi_{\upsilon_2}(s,t)$ - vector meson propagator and $F_{\upsilon_2 NN}(t)$ - form factor in the $\upsilon_2NN$ vertex.
Terms of Eq.(\ref{current}) contained in curly braces do not depend on the CM momentum in the $\pi\eta$ 
($K\overline{K}$) system thus 
they can be factorized out of the partial wave expansion. So, after all terms that do not depend on the $k_1$
or $k_2$ momenta are 
factorized out of Eq.(\ref{pwe}) we arrive at the $S$-wave projected tensor defined as
\begin{multline}
 P_{r,mm'}^{00,\mu\nu}= \frac{1}{\sqrt{4\pi}}\int d\Omega
 {Y^0_0}^\ast(\Omega)\\\times(\alpha_{r,mm'}g^{\mu\nu}+k_1^{\mu}\beta_{1r,mm'}^\nu+
k_2^\mu\beta_{2r,mm'}^\nu).
\end{multline}
The only matrix elements of the tensor
 $P_{r,mm'}^{00}$ which enter the amplitude are $P_{r,mm'}^{00,i0}$ and 
 $P_{r,mm'}^{00,ij}$,
where $i,j=x,y,z$. This results from the definition of the photon polarisation 4-vector in Eq.(\ref{polvec}).
Details of the computation of the tensor $P_{r,mm'}^{LM}$ for the $S-$wave are 
discussed in \cite{JiKaLe} and for higher partial waves in \cite{BiKa}.
\section{Final state interaction amplitudes}\label{FSI}
To describe the final state rescattering which may result in resonance creation we use the model developed in 
\cite{Kaminski:1993zb} and specialized to the isovector coupled  channel $\pi\eta$-$K\overline{K}$ case in 
\cite{Lesniak:1996qx,Furman:2002cg}. Here we collect basic definitions and results obtained with this model. 
The $S-$ wave isovector $\pi\eta$ and $K\overline{K}$ interaction is described in terms of the coupled channel 
Lippman-Schwinger equations which in momentum space read
\begin{equation}
\langle q|\hat{T}|k\rangle = \langle q|\hat{V}|k\rangle+\int\frac{d^3p}{(2\pi)^3}\langle q|\hat{V}|p\rangle
\langle p|\hat{G}|p\rangle
\langle p|\hat{T}|k\rangle,
\label{L-S}
\end{equation}
where $\hat{V}$, $\hat{G}$ and $\hat{T}$ are 2$\times$2 matrices in channel space. In this space label 
1 denotes the $\pi\eta$ channel and 2 - the $K\overline{K}$ channel. $\hat{G}$ denotes the diagonal propagator
 matrix
\begin{equation}
\langle p| G_{ij}| p \rangle=G_i(p)\delta_{ij},
\end{equation}
with the elements defined as
\begin{equation}
G_i(p)=\frac{1}{E-E_i(p)+i\epsilon}.
\end{equation}
$\hat{V}$ is the interaction matrix which is assumed in a separable form
\begin{equation}
\langle q|V_{ij}|k\rangle=\lambda_{ij}g_i(q)g_j(k),
\end{equation}
where $\lambda_{ij}$ are the coupling constants which in two channel case form a 2$\times$2 matrix
\begin{equation}
\lambda=\left(
\begin{array}{cc}
\lambda_{11}&\lambda_{12}\\
\lambda_{12}&\lambda_{22}
\end{array}
\right)
\end{equation}
and the form factor was chosen as
\begin{equation}
g_i(p)=\sqrt{\frac{4\pi}{m_i}}\frac{1}{p^2+\beta_i^2}.
\label{ff}
\end{equation}
In Eq.(\ref{ff}) $p$ is the CM momentum in $i-$th channel, $m_i$-reduced mass in this channel and $\beta_i$ 
may be interpreted as a range parameter.
Altogether the model of the final state scattering has 5 parameters - 3 coupling constants and 2 range parameters.
For the Lippmann-Schwinger equation with separable interaction we can proceed further if the amplitude $T_{ij}$ is 
rewritten in terms of reduced amplitudes $t_{ij}$
\begin{equation}
\langle q|T_{ij}|k\rangle=g_i(q)t_{ij}g_j(k),
\end{equation}
where the reduced amplitudes $t_{ij}$ depend only on the total energy.
After such substitution, the coupled integral equations Eq.(\ref{L-S}) are replaced with a set of algebraic 
equations for reduced amplitudes $t_{ij}$ which can be expressed in terms of the 2$\times$2 matrix equation
\begin{equation}
\hat{t}=\hat{\lambda}+\hat{\lambda}\hat{I}\hat{t},
\label{reducedL-S}
\end{equation}
where the nonvanishing diagonal elements of the matrix $\hat{I}$ have the form
\begin{equation}
I_{ii}=\int\frac{d^3p}{(2\pi)^3}g_i(p)G_i(p)g_i(p).
\label{integral}
\end{equation}
Equation (\ref{reducedL-S}) can be solved:
\begin{equation}
\hat{t}=(\hat{1}-\hat{\lambda}\hat{I})^{-1}\hat{\lambda},
\label{solution}
\end{equation}
where $\hat{1}$ is 2$\times$2 unit matrix.
Reduced amplitudes $t_{ij}$ defined by Eq.(\ref{solution}) are proportional to inverse of the determinant
\begin{equation}
D(E)=det(\hat{1}-\hat{\lambda}\hat{I})
\end{equation}
which defines the Jost function. The on-shell scattering matrix elements $T_{ij}(k_i,k_j)$ can be expressed in terms
of the $S$-matrix elements
\begin{equation}
S_{ij}=\delta_{ij}-\frac{i}{\pi}\sqrt{k_i\alpha_i k_j\alpha_j}\;T_{ij}(k_i,k_j)
\end{equation}
where $i\text{ and }j$ enumerate channels, $k_i$, $k_j$ are center of mass momenta in respective channels and 
$\alpha_i$ are defined as:
\begin{equation}
\alpha_1=\frac{E_\pi E_\eta}{E_\pi+E_\eta}
\end{equation}
and
\begin{equation}
\alpha_2=\frac{E_K}{2}.
\end{equation}
Here $E_{\pi}$, $E_\eta$ and $E_K$ denote the $\pi$, $\eta$ and $K$ energies in the center of mass of respective 
channels.
Jost function can be explicitly expressed in terms of the channel momenta $k_1$ and $k_2$ as
\begin{equation}
D(k_1,k_2)=D_1(k_1)D_2(k_2)-\Lambda_{12}^2 J_{11}(k_1)J_{22}(k_2)
\end{equation}
where
\begin{equation}
D_i(k_i)=1-\Lambda_{ii}J_{ii}(k_i)
\end{equation}
and $\Lambda_{ij}$ are dimensionless couplings defined as
\begin{equation}
\Lambda_{ij}=\frac{\lambda_{ij}}{2(\beta_i\beta_j)^{3/2}}.
\end{equation}
$J_{ii}$ is a redefined integral given by Eq.(\ref{integral}) and reads
\begin{equation}
J_{ii}=2 \beta_i^3 I_{ii}.
\end{equation}
For the 2 channel case, the elements of the $S-$matrix can be related to the Jost function with the following 
expressions:
\begin{equation}
S_{11}=\frac{D(-k_1,k_2)}{D(k_1,k_2)},
\label{s11}
\end{equation}
\begin{equation}
S_{22}=\frac{D(k_1,-k_2)}{D(k_1,k_2)},
\end{equation}
\begin{equation}
S_{12}^2=S_{11}S_{22}-\frac{D(-k_1,-k_2)}{D(k_1,k_2)}.
\label{s12}
\end{equation}
Given the fact that we do not have at our disposal any $\pi\eta$ scattering data, the only way to constrain model 
parameters is to exploit the information on the location of the $a_0(980)$ and $a_0(1450)$ resonances on the complex
energy plane. We denote these energies as $E^r$ and $E^R$ respectively. These are related to poles of the $S-$matrix 
elements or zeroes of the Jost function as implied by Eqs.(\ref{s11}-\ref{s12}). The complex momenta 
$k_1^r$, $k_2^r$ and $k_1^R$, $k_2^R$ are then used to obtain 2 complex (or 4 real) equations:
\begin{equation}
D(k_1^r,k_2^r)=0,
\end{equation}
\begin{equation}
D(k_1^R,k_2^R)=0.
\end{equation}
As a result one can constrain 4 parameters. To evaluate the 5-th parameter the information on the values of 
resonance branching fractions was exploited (see \cite{Furman:2002cg} for details). For further calculations we use 
the following values of the final state interaction parameters: 
$\Lambda_{11}=-0.032321$, $\beta_1=20.0$ GeV, $\Lambda_{22}=-0.068173$,
$\beta_2=21.831$ GeV and $\Lambda_{12}=5.0152\times 10^{-4}$.
\section{Numerical results}\label{Results}
\subsection{$\mathbf{S}-$wave cross sections}
\begin{figure}[htbp]
\centering
\includegraphics[scale=.35,clip]{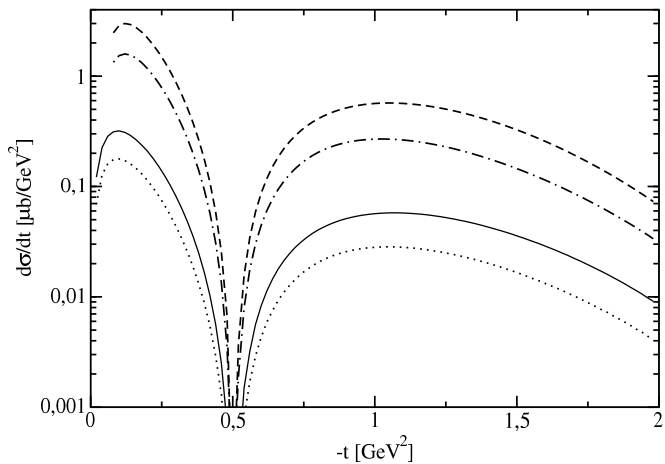}
\caption{Differential cross sections calculated at photon energy $E_\gamma=7$ GeV. Solid and dotted lines correspond 
to integration of  Eq. (\ref{doublecrosssect}) around the mass of $a_0(980)$ using complete and Born amplitudes, 
respectively. Dashed and dash-dotted line correspond to integration of  Eq. (\ref{doublecrosssect}) around the mass 
of $a_0(1450)$ using complete and Born amplitudes, respectively.}
\label{dsdt}
\end{figure}

\begin{figure}[htbp]
\centering
\includegraphics[scale=.35,clip]{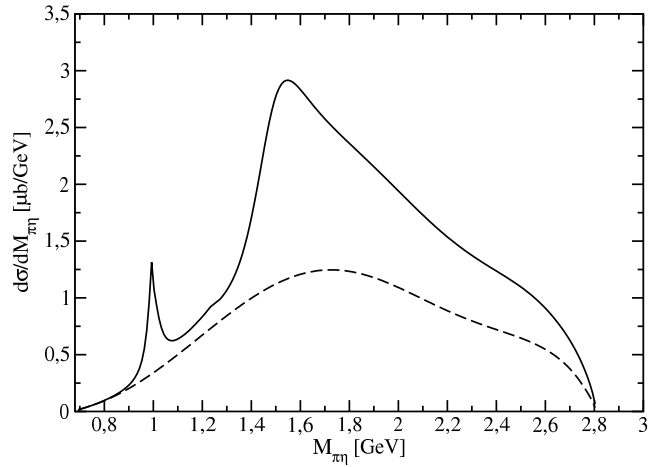}
\caption{Mass distribution $d\sigma/dM_{\pi\eta}$ at photon energy $E_\gamma$=7 GeV. Calculation based on the 
complete amplitude - solid line, and Born amplitude - dashed line.}
\label{dsdM7GeV}
\end{figure}

For the $L-$th partial wave we define the double differential cross section as
\begin{equation}
\frac{d\sigma^L}{dtdM_{\pi\eta}}=
\frac{1}{4}\frac{1}{(2\pi)^3}\frac{|k|}{32m^2{E_\gamma}^2}\sum_{\lambda_{\gamma},\lambda,M,\lambda'}
|\langle \lambda' M |A_{\pi\eta}^L|\lambda_\gamma \lambda \rangle|^2,
\label{doublecrosssect}
\end{equation}
where $m$ and $E_\gamma$ are the proton mass and photon energy in the laboratory frame respectively. For the 
calculation of the differential cross sections of the $a_0(980)$ and $a_0(1450)$ photoproduction we perform the mass 
integration of Eq.(\ref{doublecrosssect}) in the ranges determined by masses and widths of these mesons
\cite{PDG2014}. The
calculation was performed at the photon energy $E_\gamma$=7 GeV which roughly corresponds to energies
accessible in new and upgraded JLab experiments. As we are interested in the high energy regime we use the reggeised 
version of the photoproduction amplitude i.e. instead of using the normal propagator
\begin{equation}
\Pi(t)=\frac{1}{t-m_{\upsilon_2}^2}
\label{normal}
\end{equation} 
in the lower vector meson line of diagram shown in Fig. \ref{diag-fsi}, we use it's reggeised version
\begin{equation}
\Pi(s,t)=\frac{-1}{2s^{\alpha_0}}\left(1-e^{i\pi\alpha_{\upsilon_2}(t)}\right)\Gamma\left(1-\alpha_{\upsilon_2}(t)
\right)(\alpha' s)^{\alpha_{\upsilon_2}(t)},
\end{equation}
where Regge propagators of $\rho$ and $\omega$ depend on trajectories parametrized as 
$\alpha_{\upsilon_2}(t)=\alpha_0+\alpha'
(t-m_{\upsilon_2}^2)$, with $\alpha_0$=1.0 and $\alpha'$=0.9 GeV$^{-2}$ and $m_{\upsilon_2}$ being the mass of the 
$\rho$ or $\omega$
respectively. The cross sections calculated with both complete amplitudes and Born amplitudes are shown in Fig. \ref{dsdt}.
Comparing the Born cross sections and complete cross sections, one sees that the combined effect of the final state 
interactions and interchannel coupling in $M_{\pi\eta}$ regions corresponding to $a_0(980)$ and $a_0(1450)$ is 
to increase the cross section by a factor of about 2. The result we obtained for $a_0(980)$ is compatible with 
that obtained in \cite{DonKal}, where $a_0(980)$ was assumed to be a member of a ground state $q\overline{q}$ 
scalar nonet.
On the other hand our cross section for $a_0(1450)$ is larger than the corresponding value based on the quark model
by about an order of magnitude.
Note, however, that the solid and dashed curves in Fig. 
\ref{dsdt} describe not just the $a_0(980)$ and $a_0(1450)$ photoproduction cross sections but rather these cross 
sections multiplied by the branching fraction to the $\pi\eta$ channel.
The minimum observed at $t$=-0.5 GeV$^2$ is characteristic to models with reggeon exchange. In 
applications, additional contributions, like Regge cuts \cite{Donnachie:2015jaa} are used to "fill" that 
minimum.
For the same photon energy we calculated the mass distribution which is shown in Fig. \ref{dsdM7GeV}.
Here we see pronounced signals of both $a_0(980)$ and $a_0(1450)$ albeit both on considerable Born background. 
We stress, however, that the status of our calculations for $a_0(980)$ and $a_0(1450)$ is rather different. In case 
of $a_0(980)$ whose molecular component may be dominant, the obtained mass distribution can be treated as true 
prediction. The strong signal obtained for $a_0(1450)$ can not be confronted with experimental data at present but 
seems unrealistic. Thus it may be interpreted as indication 
that this resonance is rather a member of the $q\overline{q}$ nonet than the dynamically created state. 
In such case description of this resonance would be beyond the area of applicability of our
model. 
Another possible explanation is that the amplitude used to describe the final state interactions does not take into 
account the $\pi\eta'$ decay channel of the $a_0(1450)$. This in turn may result in overestimation of the 
strength of photoproduction amplitude in the $\pi\eta$ channel.

\begin{figure}[htbp]
\centering
\includegraphics[scale=.35,clip]{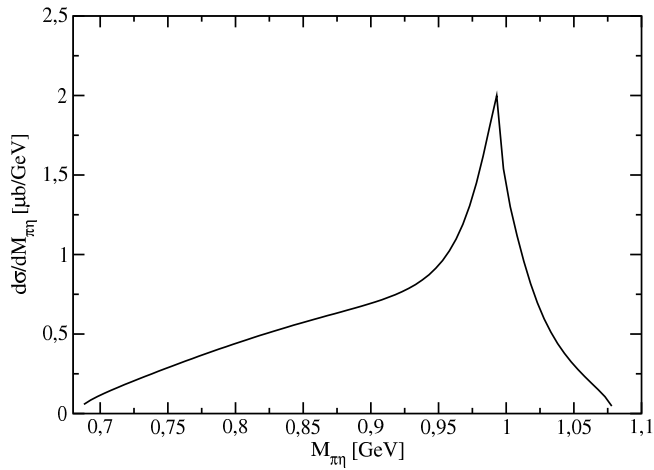}
\caption{Mass distribution $d\sigma/dM_{\pi\eta}$ at photon energy $E_\gamma$=1.7 GeV.}
\label{dsdm1.7}
\end{figure}

In order to confront our predictions with other models at lower energies \cite{MaOsTo,DonKal}, 
we calculated the mass distribution at 
the photon energy of 1.7 GeV which is close to energy used TAPS/Crystal Ball at Mainz and CB-ELSA, see Fig. 
\ref{dsdm1.7}. For this calculation we used normal propagators in the photoproduction amplitude (Eq. (\ref{normal})).
Our low energy predictions are compatible with calculations based on chiral unitary model \cite{MaOsTo} and on quark 
model \cite{DonKal}, although the latter calculation was performed at higher photon energy of 5 GeV. It is 
remarkable, however, that the mass distribution in the region of $a_0(980)$ we obtain, is about 2 orders
of magnitude larger than the one obtained by Donnachie and Kalashikova based on pseudoscalar loop model. According
to our calculation, the photoproduction of scalar isovecotor resonances through dynamical formation in the final 
state makes the measurement of the $a_0(980)$ feasible at experimental conditions provided by new experiments at 
Jefferson Laboratory. Similar predictions by other models make it inevitable to discriminate among the models by application of the more stringent tests based on the analysis of the polarization data.
\subsection{Higher partial waves}
\begin{figure}[htbp]
\centering
\includegraphics[scale=.35,clip]{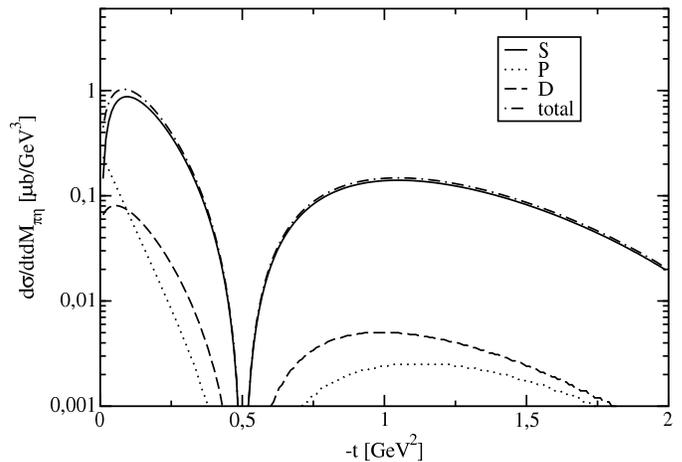}
\caption{Born $\pi\eta$ photoproduction cross section for partial waves $S$, $P$ and $D$ at photon energy 
$E_\gamma$=7 GeV and $M_{\pi\eta}$ corresponding to $a_0(980)$.}
\label{a0_980}
\end{figure}

\begin{figure}[htbp]
\centering
\includegraphics[scale=.35,clip]{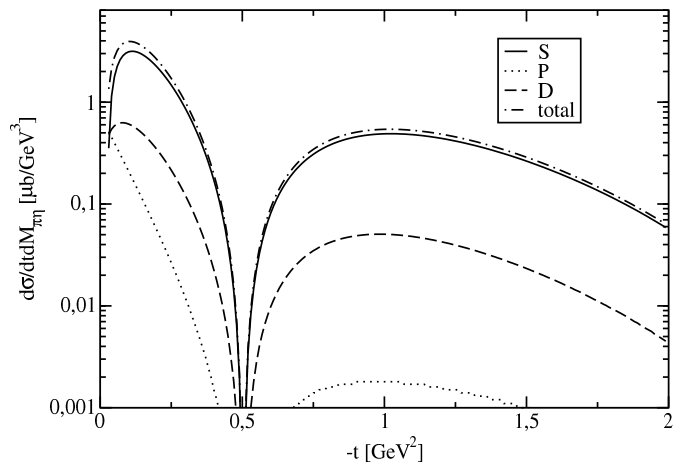}
\caption{Born $\pi\eta$ photoproduction cross section for partial waves $S$, $P$ and $D$ at photon energy 
$E_\gamma$=7 GeV and $M_{\pi\eta}$ corresponding to $a_0(1450)$.}
\label{a0_1450}
\end{figure}

As seen from Fig. \ref{dsdM7GeV} the Born component accounts for the large part of the overall $S-$wave cross 
section. It is then natural to expect that this property should hold also for higher partial waves where we don't 
have the final state rescattering amplitudes at our disposal. For these waves we can treat the 
Born cross sections as crude estimations of cross sections in the full model.  
Thus, we calculate the Born cross sections for the partial waves $S$, $P$, $D$ and the 
sum of all partial waves at the 
fixed $M_{\pi\eta}$ masses corresponding to resonances $a_0(980)$ (Fig.\ref{a0_980}) and $a_0(1450)$ 
(Fig.\ref{a0_1450}). The calculation was again performed at photon energy $E_\gamma$=7 GeV and the reggeised version 
of the amplitude was used.
We see that apart from the very forward region the double differential cross section is dominated by the 
$S-$wave. Moreover, the lowest partial waves practically saturate the Born cross section (we do not show the 
sum of the $S$, $P$ and $D-$waves as it is practically indistinguishable from the full cross 
section). The clear suppression of the odd partial waves results from the cancellation of the terms in the amplitude
which are odd functions of the angles $\theta$ and $\phi$ which describe the direction of $\pi$ in the 
helicity system. Thus for masses around 1.5 GeV and outside the forward region the $P-$ wave is over an order of 
magnitude smaller than $D-$wave and over two orders of magnitude smaller than the $S-$wave. For masses around 1 GeV
the $P-$wave suppression is also visible, although for the very small values of $-t$ the $P-$wave dominates 
which results from the fact that the angular momentum of $\pi\eta$ system "inherits" the helicity of the incident
photon.
\section{Summary and outlook}\label{Summary}
Assuming that isovector resonances $a_0(980)$ and $a_0(1450)$ are dynamically produced in the final state we 
calculated the mass spectra
at both the low photon energies and at higher energies accessible in upgraded JLab facilities. 
The model includes both the pseudoscalars and vectors in the intermediate meson loop. This is why we do not observe
the strong amplitude suppression characteristic to models which include only pseudoscalar loops.
Our predictions for $a_0(980)$ at photon energy $E_\gamma=$1.7 GeV
are compatible with calculations performed with quark model and chiral unitary model. This makes the values of the $a_0(980)$ photoproduction cross section large enough to be measured. 

The situation with $a_0(1450)$ is quite 
different. Molecular nature of this resonance is disputable. Moreover, the mass of the $a_0(1450)$ is above the 
$\pi\eta'$ threshold, so that inclusion of the third channel, may be necessary to properly describe the $a_0(1450)$
photoproduction. Taking the $\pi\eta'$ channel into account may affect the inter-channel couplings and decrease the 
strength of both $\pi\eta$ elastic amplitude and $K\overline{K}\to\pi\eta$ transition amplitude.
In this introductory analysis we neglected the effects of the off-shell meson propagation in the intermediate state,
which means that apart from the coupling constants, and form factor parameters drawn from other models our 
approach does not engage any new parameters. 
As mass distributions obtained at low photon energies in different production mechanisms are rather compatible there 
is the need for the analysis based on polarization data to make the more stringent discrimination among the models.
We also calculated the Born cross sections for partial waves $P$ and $D$. These can be treated as crude estimations 
of the cross sections to be computed when reliable parameterisations of the final state interactions in higher 
partial waves become available. Having compared the Born cross sections for lowest partial waves we conclude that 
the $S-$wave dominates in the $\pi\eta$ channel and the odd partial waves are suppressed with respect to the even 
ones.
\begin{acknowledgments}
This research has been funded by the Polish National Science Center (NCN) grant No. DEC-2013/09/B/ST2/04382
\end{acknowledgments}

\end{document}